\documentclass[12pt]{iopart}
\usepackage{latexsym}
\usepackage{graphicx}
\include{biblio}
\def\newblock{\hskip .11em plus .33em minus .07em}

\begin{document}

\title[Probing the birth of attosecond pulses using a two-color field]{Quantum mechanical approach to probing the birth of attosecond pulses using a two-color field}
\author{J. M. Dahlstr\"om, A. L'Huillier and J. Mauritsson} 
\address{Department of Physics, Lund University, P.O. Box 118, SE-221 00 Lund, Sweden}
\ead{marcus.dahlstrom@fysik.lth.se}
\begin{abstract}
We investigate the generation of even and odd harmonics using an intense laser and a weak second harmonic field. Our theoretical approach is based on solving the saddle-point equations within the Strong Field Approximation. 
The phase of the even harmonic oscillation as a function of the delay between fundamental and second harmonic field is calculated and its variation with energy is found to be in good agreement with recent experimental results. We also find that the relationship between this phase variation and the group delay of the attosecond pulses, depends on the intensity and wavelength of the fundamental field as well as the ionization potential of the atom. 

\end{abstract}

\maketitle

\section{Introduction}

Attosecond pulses \cite{PaulScience2001,HentschelNature2001} are created through the
interaction between intense infrared (IR) laser fields and atoms or molecules in a
process known as high-order harmonic generation (HHG). This process is initiated by the creation of an electron wave packet through tunneling ionization, followed by acceleration in the laser field and
recombination with the ion core resulting in extreme ultraviolet (XUV)
emission confined to a fraction of the laser cycle. The generation process can be controlled by shaping the driving laser field, \textit{e.g.} by coherently adding laser fields of different wavelengths to the fundamental laser field \cite{LongPRA95,GaardePRA1996,RadnorPRA2008,ChipperfieldPRL2009}. Most investigations have been carried out by adding the second harmonic generated in a doubling crystal \cite{FigueiraPRA2000,MauritssonPRL2006,ManstenNJP2008,ishiiOE2008,dudovichPRA2009,FrolovPRA2010,HePRA2010,EilanlouOX2010}, with various goals ranging from optimization of the conversion efficiency, characterization of the emitted attosecond pulses and production of attosecond pulse trains with one pulse per cycle of the fundamental field. Several parameters can be varied in these experiments: the intensity ratio, the phase difference and the relative polarization direction.  

\begin{figure}[ht]
	\centering
		\includegraphics[width=0.75\linewidth]{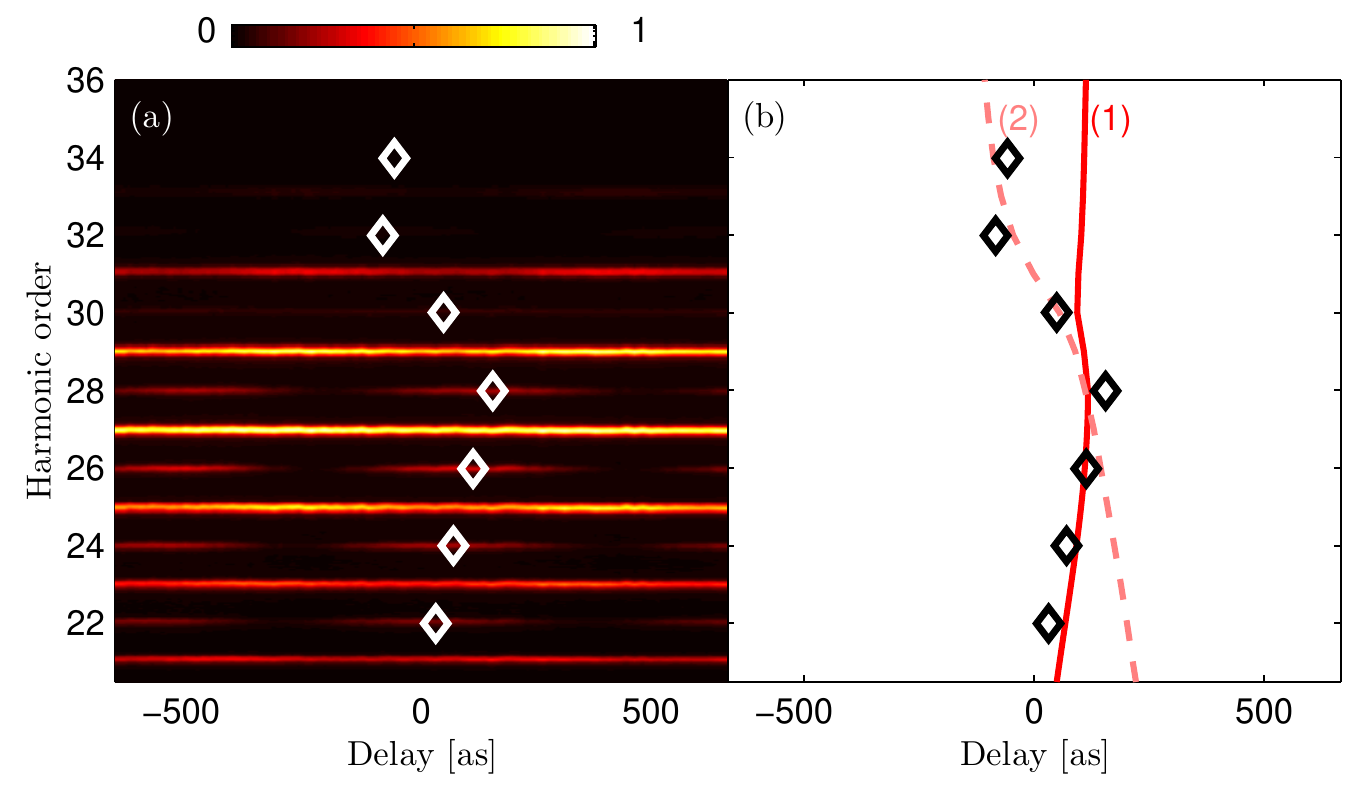}
		\caption{(a) Experimental two-color high-order harmonic spectrogram over relative delay \cite{HePRA2010}. The diamonds ($\diamond$) indicate the experimental delays corresponding to maxima of the modulation in the even harmonic orders. (b) Calculated delays corresponding to maxima of the even orders for an effective intensity of $1.4\times10^{14}$ W/cm$^2$. Full (dashed) curve are calculated for the short (long) branch labeled 1 (2). Below (above) harmonic order 28 the short (long) branch fits the experimental data ($\diamond$).}
\label{first}
\end{figure}

When the second harmonic field is a weak perturbation to the fundamental,  
the generation of odd harmonics is barely changed, but the induced symmetry breaking leads to the appearance of 
weak even harmonics. 
High-order harmonic spectra recorded as a function of delay between the two fields show that the intensity of the even harmonics is modulated and that this modulation has an offset depending on the harmonic order, as shown in Figure \ref{first}(a) \cite{HePRA2010}. 
These photon spectrograms are at a first glance similar to the photoelectron 
spectrograms of the RABITT (Reconstruction of Attosecond Bursts by Interference of Two-photon Transitions) method \cite{PaulScience2001,MairesseScience2003} 
used to characterize attosecond pulses. Dudovich \textit{et al.} \cite{DudovichNP2006} suggested that 
 the two-color HHG spectrograms could be used to determine the emission times of the 
attosecond pulses \textsl{in situ}, thus ``probing the birth of attosecond pulses''. This method was then applied by Doumy \textit{et. al.} \cite{DoumyPRL2009}, using laser systems of different wavelengths, and by us \cite{DahlstromPRA2009}, in a direct experimental comparison with the RABITT method. 

In this article we investigate HHG by a laser field and its second harmonic by solving the Schr\"{o}dinger equation within the Strong Field Approximation (SFA), using the saddle point equation method \cite{LewensteinPRA1995}. We assume throughout that the second harmonic is weak and can be considered as a small perturbation to the fundamental field. We calculate the phase variation of the even harmonic oscillation and compare it with the harmonic emission times. Our model allows us also to interpret the rapid variation of this phase at high energy, observed in several experiments \cite{DahlstromPRA2009,HePRA2010}, as
a change of the dominant quasiclassical trajectory from the short to the long one [Figure \ref{first}(b)].
The article is organized as follows: In section 2 we review and compare classical and quasiclassical electron trajectories for one-color (Sec. 2.1) and two-color HHG (Sec. 2.2). In section 3 we study the generation of even harmonics close to the cutoff (Sec. 3.1) and we relate the phase of the even harmonic oscillation to the emission times of the attosecond pulses (Sec. 3.2). In section 4 we summarize our results by presenting a more general relationship between the phase of the even harmonic oscillation and the emission times. 

%


\section{Quasiclassical trajectories}

Our method is based on the stationary phase method (also referred to as the saddle point method) to solve the Schr\"{o}dinger equation for an atom exposed to a strong laser field using the SFA \cite{LewensteinPRA1994}.
In particular, we consider the interaction of an atom with a strong IR laser field (frequency $\omega$) and a weak second harmonic (frequency $2\omega$). 
In the SFA, the electron will (I) tunnel into the continuum, (II) accelerate in the strong laser field and then (III) return to the atom and emit a harmonic photon (frequency $\Omega$). The phase of the harmonic radiation is related to the quasiclassical action of the electron
\begin{equation}
S(\vec{p},t,t_0)=\int_{t_0}^{t} dt'\left(\frac{(\vec{p}-e\vec{A}(t'))^2}{2m}+I_p\right),
\label{action}
\end{equation}
where $\vec{A}(t)$ is the vector potential of the laser field, $I_p$ is the ionization potential of the atom; and $\vec{p}$, $t_0$, $t$, $m$ and $e$ are the drift momentum, tunneling time, return time, mass and charge of the electron respectively.
The high-order harmonic emission will mainly originate from the stationary points of $S-\hbar\Omega t$, 
 with respect to all variables $\left[\vec{p},t,t_0\right]$, which satisfy the following three equations \cite{LewensteinPRA1994}
%
%
%
\begin{equation}
\int_{t_0}^{t}dt'~e\vec{A}(t') = (t-t_0)\vec{p}, 
\label{eqn1} 
\end{equation}
\begin{equation}
\frac{(\vec{p}-e\vec{A}(t_0))^2}{2m}=-I_p, 
\label{eqn2} 
\end{equation}
\begin{equation}
\frac{(\vec{p}-e\vec{A}(t))^2}{2m}=\hbar\Omega-I_p.
\label{eqn3}
\end{equation}
The electron is thus required to return to the atom at time $t$ [Eq.~\ref{eqn1}], to undergo complex tunneling at time $t_0$ [Eq.~\ref{eqn2}] and to satisfy energy conservation [Eq.~\ref{eqn3}]. 
For any realistic atom, we have $I_p>0$ which implies that the electron must tunnel into the continuum.
The tunneling process results in damping of the electron trajectories, \textit{i.e.} 
complex stationary points in the harmonic plateau \cite{BeckerAAMO2002}. 
The electron trajectories beyond the cutoff are always 
strongly damped (also for $I_p=0$), because they are always classically forbidden,
corresponding to large imaginary components of the stationary points.
The physical emission time of a given harmonic is given by the real part of the complex emission time \cite{MairesseScience2003,ChirilaPRA2010,VarjuJMO2005}. 
\subsection{One-color case}
\begin{figure}[ht]
	\centering
		\includegraphics[width=0.75\linewidth]{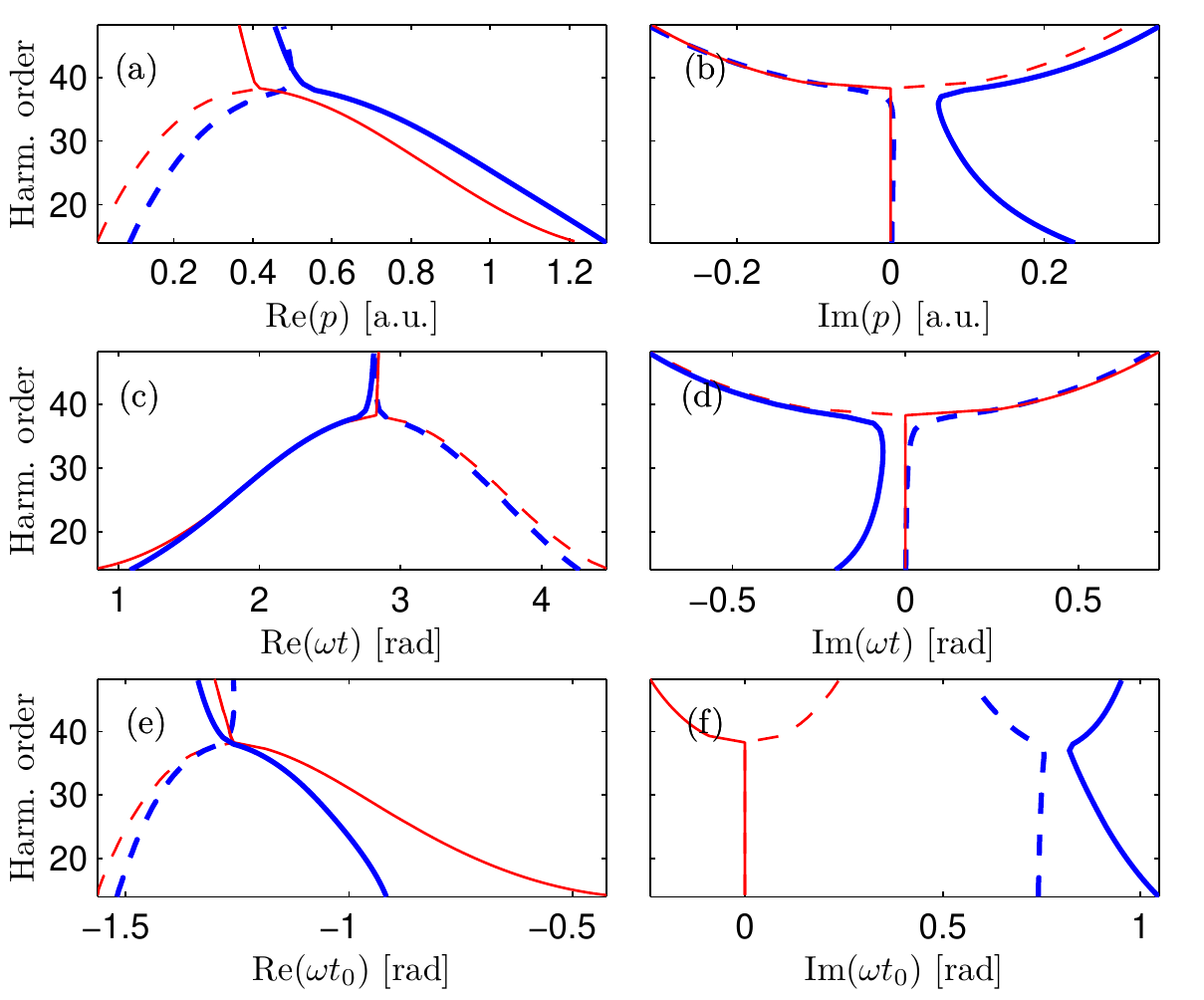}
\caption{Real and imaginary parts of the stationary points, $[p,t,t_0]$. Ionization potentials:  argon $I_p^{(Ar)}=15.76$ eV (thick blue) and classical $I_p^{(0)}=0$ eV (thin red). The classical case is shifted by $1.3I_p$ to match the quantum mechanical cutoff. The short branch is a line; while the long branch is a dashed line. The intensity is $2\times10^{14}$ W/cm$^2$ and the wavelength is 800 nm.}
\label{points}
\end{figure}

We first consider the one-color case
with a vector potential, $\vec{A}(t)=\vec{A}_1\sin(\omega t)$, where $\omega$ is the angular frequency of the fundamental laser light.
The complex stationary points $[\vec{p}^{(n)},t^{(n)},t_0^{(n)}]$ are calculated as a function of the high-order harmonic photon angular frequency, $\Omega$. The index $n$ is used to separate different sets of solutions, where $n=1$ corresponds to the short branch; and $n=2$ corresponds to the long branch. 
%
%
%
A direct comparison between the stationary points for $I_p=15.76$ eV (Ar) and $I_p=0$ eV (referred to as the classical case) is shown in Figure \ref{points}. 
The detailed behavior of these stationary points will prove to be important not only for the generation of attosecond pulses from one-color HHG, but also in the quantitative analysis of two-color HHG.
The classical case corresponds to solving Eqs.~\ref{eqn1}-\ref{eqn3} for $I_p=0$. 
The stationary points are first real, describing classical trajectories in the continuum. Using classical trajectories to explain the HHG process leads to the approximate cutoff law: $\hbar\Omega_{max} \approx 3.2U_p + I_p$ \cite{KrausePRL1992,CorkumPRL1993}. 
Harmonics above this cutoff can not be generated because there are no electrons returning with suffiecient kinetic energy.
Quantum mechanically, we see that the stationary points quickly develop large imaginary components above the cutoff, which result in an exponential damping of the harmonic yield. 
Solving Eqs.~\ref{eqn1}-\ref{eqn3} for $I_p \neq 0$, the exact cutoff law is found to be $\hbar\Omega_{max}=1.3I_p+3.2U_p$ \cite{LewensteinPRA1994}. 
In Figure \ref{points} we have, therefore, shifted the classical solutions by $1.3I_p$ so that the cutoff harmonic of the classical case coincides with the quantum mechanical case of argon for a more meaningful comparison. 
The similarities and differences between the classical case (red line) and the quantum case (blue line) are clearly observed: The real parts of the drift momenta [Fig.\ref{points}(a)] are almost the same in the two cases, apart from a small systematic shift. The real parts of the return times [Fig.\ref{points}(c)] overlap almost perfectly in the main part of the plateau for both the short (line) and long (dashed) branches. This demonstrates the usefulness of the classical model for understanding the intrinsic chirp of the returning electron wave packet and the resulting chirp of the attosecond light pulses. 
The real parts of the tunneling times [Fig.\ref{points}(e)] are, however, quite different. In the lower part of the plateau there is a discrepancy of a factor of two for the short branch. The short branch trajectories tunnel at earlier times compared to the classical case due to the ionization potential. This implies longer times in the continuum and an increased amount of phase acquired, as the path of integration extends in Equation \ref{action}. For the long branch the trend is opposite and smaller.
The imaginary parts of the stationary points [Fig.\ref{points}(b,d,f)] lead to exponential damping of the electron amplitudes. This is especially clear for the imaginary part of the tunneling time [Fig.\ref{points}(f)] where the complex part is zero in the classical case where no tunneling is required; while, in the complex case, the short branch will suffer more damping than the long branch due to a lower instantaneous electric field strength at the time of tunneling.

The high-order harmonic dipole from one \textit{half} period of the laser field can be approximated as a sum of stationary contributions 
\begin{equation}
\vec{x}_{\Omega} = 
\int d^3p \int_0^{T/2} dt \int_{-\infty}^t dt_0 
\Lambda
\exp \left[ i S(\vec{p},t,t_0)/\hbar - i\Omega t \right]
\nonumber 
\approx \sum_{n=1,2} \vec{x}_{\Omega}^{(n)}.
\end{equation}
These discrete contributions correspond to different sets of stationary phase solutions
\begin{equation}
\vec{x}_\Omega^{(n)}=
\Lambda^{(n)}
\left(
\frac{i h^5}
{\det[M_{ij}^{(n)}]}
\right)^{1/2}
\exp \left[ i S^{(n)}/\hbar - i \Omega t^{(n)} \right],
\label{reconstruct}
\end{equation}
where $S^{(n)}=S(\vec{p}^{(n)},t^{(n)},t_0^{(n)})$ and where 
\begin{equation}
 M_{ij} = 
\frac{\partial^2}{\partial i \partial j} \left[ S(\vec{p},t,t_0) - \hbar\Omega t \right],
\label{matrix}
\end{equation} 
is the Hessian matrix of second-order derivatives of the Legendre transformed action with $i$ and $j$ being any of our variables $[\vec{p},t,t_0]$.
The Hessian matrix is complex symmetric and we have assumed that it can be diagonalized so that the one-dimensional stationary phase approximation can be applied to all five integrals independently. 
The prefactor $\Lambda = d(\vec{p}-e\vec{A}(t))^{*}E(t_0)d(\vec{p}-e\vec{A}(t_0))$, describes the dipole transitions to the continuum at time $t_0$, and the subsequent recombination to the ground state at time $t$.
In the following, we will assume that $\Lambda$ is slowly varying and that it, therefore, does not affect the stationary points. 
A more exact analysis could, however, also include the effect of the energy dependent atomic scattering phases, $\eta_l$, in the stationary phase equations (Eqs.~\ref{eqn1}-\ref{eqn3})
(from the recombination matrix element $d(\vec{p}-e\vec{A}(t))^{*} \propto \exp[-i\eta_l]$).
This would lead to a small change in the stationary points, 
which would be especially interesting to study in argon due 
to an unusually strong variation of the scattering phase 
in the present energy region \cite{KennedyPRA1972}.
%

Assuming the process to be periodic, the total dipole response, $\vec{X}_\Omega$, is found by summing the stationary points from one \textit{whole} period of the fundamental laser field, $0<t<T$, \textit{i.e.} two adjacent half periods. There are thus two contributions from each branch, 
\begin{equation}
\vec{x}^{(n)}_{\Omega}(t+T/2,t_0+T/2) = 
-\vec{x}^{(n)}_{\Omega}(t,t_0)  
\exp\left[-i\Omega T/2 \right],
\label{halfperiods}
\end{equation}
separated by a half period, $T/2$. The overall minus sign comes from the change of sign of $E(t)$; while the phase factor originates from the Fourier component in the Legendre transformed action.
The total dipole response becomes 
\begin{equation}
\vec{X}_{\Omega}  \approx  \sum_{n} \vec{x}_{\Omega}^{(n)}
\left\{
1-\exp\left[ -i \Omega T/2 \right]
\right\}
= 
2\sum_{n} \vec{x}_{\Omega}^{(n)} \times
\left\{
\begin{array}{ll}
1, &  \Omega/\omega \  \mbox{is odd}
\\ 
0, &  \Omega/\omega \  \mbox{is even}\ 
\end{array}
\right. , 
\label{onecolordipole}
\end{equation}
where the two contributions add constructively for odd orders; and destructively cancel for even orders.

 
 
\subsection{Two-color case}

We now consider a two-color laser field composed of a fundamental laser field and a weak second harmonic with the same polarization, $\vec{A}(t)=\vec{A}_1\sin(\omega t)+\lambda\vec{A}_2\sin(2\omega t +\phi)$, where $\lambda$ is a small perturbation parameter. 
%
%
%
It is possible to solve the two-color high-order harmonic emission using the stationary phase equations [Eqs.~\ref{eqn1}-\ref{eqn3}] directly, but this requires evaluation of the system at all values of $\phi$. We will follow a different route where the second harmonic is treated as a perturbation and only the stationary points of one-color HHG need to be calculated. The two-color action is expanded in $\lambda$ as
\begin{equation}
S \approx 
\underbrace{\int_{t_0}^{t} dt'\left(\frac{[\vec{p}-e\vec{A}_1(t')]^2}{2m}+I_p\right)}_{S_1} 
\underbrace{-\lambda \int_{t_0}^{t}dt'\frac{[\vec{p}-e\vec{A}_1(t')][e\vec{A}_2(t',\phi)]}{m}}_{\lambda\sigma}.
\label{action2}
\end{equation}
The first term in Eq.~\ref{action2} corresponds to the action in the one-color case, $S_1$; and the second term is the correction term due to the interaction with the second harmonic field, $\sigma=\sigma(\phi)$. The correction term can be calculated as
\begin{equation}
\sigma = 
\frac{e}{m}\left[
\vec{p}\vec{A}_2\frac{\cos(2\omega t'+\phi)}{2\omega} + \right.
\left. e\vec{A_1}\vec{A_2}\left( \frac{\sin(3\omega t'+\phi) }{6\omega}  - \frac{\sin(\omega t'+\phi)}{2\omega} \right)
\right]_{t_0}^{t},
\label{sigma}
\end{equation}
%
%
where  
$\sigma$ depends on the ionization potential through the stationary points. 
Note that the derivation of Eq.~\ref{sigma} using the quantum mechanical stationary points differs from those presented previously \cite{DudovichNP2006,DahlstromPRA2009}, because it includes the effect of the ionization potential within the SFA. 

The two-color high-order harmonic dipole can be calculated by using Eq.~(\ref{reconstruct}) and by adding an additional slow factor due to the second harmonic. In analogy with Eq.~(\ref{onecolordipole}), the dipole from one period of the fundamental field contains one discrete contribution for each branch and half period,

\begin{eqnarray}
\vec{X}_{\Omega} & \approx & \sum_{n} \vec{x}_{\Omega}^{(n)}
\left\{
\exp\left[i\sigma^{(n)}\right] 
-\exp\left[-i\sigma^{(n)} -\frac{i \Omega T} {2\hbar} \right]
\right\}
\nonumber
\\
&=& 
2\sum_{n} \vec{x}_{\Omega}^{(n)} \times
\left\{
\begin{array}{rl}
\cos\left[\sigma^{(n)}/\hbar\right], &  \Omega/\omega \  \mbox{is odd}
\\ 
i\sin\left[\sigma^{(n)}/\hbar\right], &  \Omega/\omega \  \mbox{is even}\ 
\end{array}
\right. 
\label{twocolordipole}
\end{eqnarray}
where $\vec{x}_\Omega^{(n)}$ is the half period contribution in the one-color field, and where the following property  
\begin{equation}
\sigma^{(n)}(t+T/2,t_0+T/2) = 
-\sigma^{(n)}(t,t_0)
\label{halfperiodssigma}
\end{equation}
relating the two-color phase between adjacent half periods for a given branch $n$ is used. 
The intensity of the even order harmonic emission from a single atom can be approximated by
\begin{equation}
I_{\Omega} \propto 
\left| 
\vec{X}_{\Omega} 
\right|^2 
\approx  
\left| 
2\sum_n \vec{x}_{\Omega}^{(n)} 
i \sigma^{(n)}/\hbar
\right|^2,
\label{evenintensityalln}
\end{equation}
which is valid for $|\sigma^{(n)}|\ll \pi$. 
We define the intensity from a specific branch, $n$, as 
\begin{equation}
I_{\Omega}^{(n)}(\phi) 
\propto  
\left| 
\sigma^{(n)}(\phi)
\right|^2. 
\label{evenintensity}
\end{equation}
This (artificial) separation of the branches can be realized in macroscopic medium either by phase matching in a long gas cell \cite{HePRA2010} or by spatial separation in the far field \cite{dudovichPRA2009,DudovichNP2006}.

\section{Subcycle delay dependence of even harmonics}

\begin{figure}[ht]
	\centering
		\includegraphics[width=0.75\linewidth]{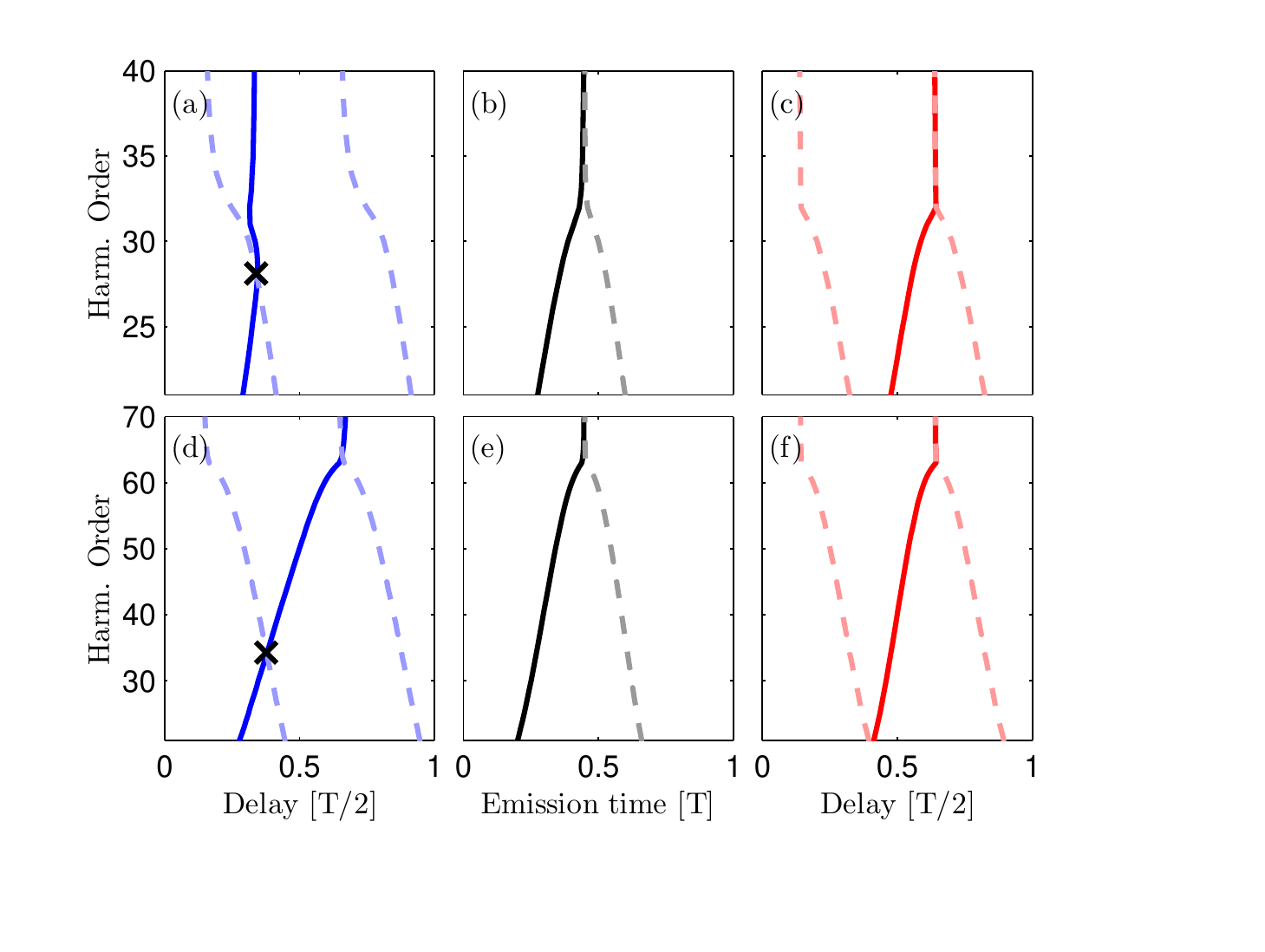}
	\caption{
Quantum mechanical in situ phases, $\phi_0^{(n)}$, between the fundamental and weak second harmonic field that maximize the even  harmonic emission in argon ($I_p^{(Ar)}=15.75$ eV) for intensities of (a) $1.5\times 10^{14}$ W/cm$^2$ and (d) $I=4\times 10^{14}$ W/cm$^2$. The in situ phases are presented as delays, $\tau_0^{(n)}=-\phi_0^{(n)}/2\omega$ in units of $T/2$.  
The corresponding high-order harmonic emission times are shown in (b) and (e), in units of $T$. 
Classical in situ phases ($I_p^{(0)}=0$) for intensities of (c) $1.5\times 10^{14}$ W/cm$^2$ and (f) $I=4\times 10^{14}$ W/cm$^2$.	
The short (long) branch is a full (dashed) curve.
$T=2\pi/\omega$ is the period of the fundamental laser field.}
\label{insituemission}
\end{figure}

One fascinating aspect of two-color HHG is that it depends on the subcycle delay (or relative phase $\phi$) between the laser fields. 
Dudovich \textit{et al.} \cite{DudovichNP2006} have proposed to use the $\phi$-dependence of even-order harmonics in order to estimate the emission times of the attosecond pulses in situ. 
%
We will refer to the relative phases that maximize the intensity of the even harmonics as the \textit{in situ phases}: $\phi_0^{(1)}(\Omega)$ and $\phi_0^{(2)}(\Omega)$, for the short and long branch respectively.
The in situ phases are plotted in Figure~\ref{insituemission}~(a) and (d) as a function of harmonic order for argon at two different IR intensities.
The corresponding emission times of the high-order harmonics, $Re\{t^{(n)}(\Omega)\}=t^{(n)}_R(\Omega)$, are shown in (b) and (e).
The intensity in (a,b) is realistic for typical laser pulses of 30 fs in argon; while the intensity in (d,e) is greater than the saturation intensity and, therefore, not experimentally feasible for multicycle laser pulses. 
The aim of our analysis is to distinguish between what can be measured experimentally, $\phi_0^{(n)}(\Omega)$, and  the desired emission times, $t^{(n)}_R(\Omega)$.
The in situ phases from the classical model ($I_p=0$) are plotted in Figure~\ref{insituemission}~(c) and (f) for comparison.  

\subsection{Behavior close to cutoff}
The similarity between the in situ phases and the emission times is striking, especially for the high intensity [Fig.~\ref{insituemission}~(d,e,f)] where the short and long branch merge in the cutoff at harmonic 63. 
At lower intensity [Fig.~\ref{insituemission}~(a)] the in situ phases do not merge in the cutoff. 
Intuitively, one might think that the short and long branch should merge in the cutoff, but this is not necessary in the quantum mechanical case. 
The stationary points of the short and long branch do \textit{not} merge on the imaginary axis in the cutoff [Fig.~\ref{points}~(b,d,f)]. 
The different behavior of the in situ phases for the short and long branch in the cutoff is, hence, an amplitude effect rather than a pure phase effect. This is verified by inserting only the real part of the stationary points into Eq.~(\ref{evenintensity}) which does indeed yield coincidental cutoff behavior of both branches, as expected for pure phase effects. 
Furthermore, it is the long branch that remains physical beyond the cutoff; while the strange behavior of the short branch arises from a set of stationary points that become unphysical beyond the cutoff \cite{LewensteinPRA1995b}.
In the classical model [Fig.~\ref{insituemission}~(c,f)], the in situ phases always merge in the cutoff since there are no amplitude effects (damping) in the plateau [Fig.~\ref{points}~(b,d,f)].

It is also worth to note that the in situ phases of the two branches intersect at lower harmonic orders than the cutoff. This ``intra-plateau crossing'' is marked with a cross ($\times$) in Figure~\ref{insituemission}~(a,d) and it should \textit{not} be confused as the position of the cutoff. At high intensities it is easy to distinguish between the intra-plateau crossing and the cutoff; while at low intensities they may be separated by a few harmonic orders only.
In Figure~\ref{first} we present an experimental result where $\phi_0$ is linear from harmonic 22 to harmonic 28. At higher harmonic orders, a dramatic bend is observed \cite{HePRA2010,DahlstromPRA2009}. Using our quantum mechanical model we identify the lower orders as part of the short branch, $n=1$, (full curve); while the higher-orders are identified as the long branch, $n=2$, (dashed curve). 
We stress that the dominance of the long branch close to the cutoff is a new result which appears already at the single atom level when the short branch becomes unphysical. This new finding illustrates the usefulness of quantitative probing of HHG using a perturbative field. This effect would be very difficult to observe with the RABITT method because the corresponding emission times always merge in the cutoff, see Fig.~\ref{insituemission} (b,e).
The classical model fails to reproduce the bend [Fig.~\ref{insituemission}~(c)] because the corresponding intra-plateau crossing occurs at much lower harmonic orders (not shown).
Having discussed the details of the behavior close to the cutoff, we now turn our attention to the central part of the harmonic plateau. 

%
%
%

\subsection{Ratio of in situ phases and emission times}

In the following, we consider the first spectral derivative of the in situ phase, $\partial\phi^{(n)}_0/\partial\Omega$, which we wish to compare with the group delay dispersion (GDD), $\partial t^{(n)}_R/ \partial\Omega$. 
%
%
Both quantities are evaluated in the central 50\% of the harmonic plateau, \textit{i.e.} in a region set by $1.3I_p + 3.2U_p \pm 0.25\times 3.2U_p$, where both $\phi_0^{(n)}(\Omega)$ and $t^{(n)}_R$ are linear to a very good approximation.
We avoid fast and nonlinear variations both close to the ionization potential ($\hbar\Omega=I_p$) and close to the cutoff ($\hbar\Omega=1.3I_p+3.2U_p$) using this central region.
We define the ratio between the two quantities as 
\begin{equation}
\gamma^{(n)} = 
-\omega\frac{\partial t^{(n)}_R}{\partial\Omega} / \frac{\partial \phi_0^{(n)}}{\partial \Omega}=
-\omega\frac{\partial t^{(n)}_R}{\partial \phi_0^{(n)}}.
\label{conversion}
\end{equation}
The GDD can in principle be obtained from the $\phi$-dependence of the even-order harmonics as 
$\partial t_R^{(n)}/\partial \omega=-\gamma^{(n)} \partial \phi^{(n)}_0/\partial\omega/\omega$. This relation would be  very useful if $\gamma^{(n)}$ was a constant (or at least a constant for each branch $n$). 
Unfortunately we show that $\gamma^{(n)}$ depends on the ionization potential of the atom, as well as the laser intensity and wavelength.

%
%
%
%
%

\subsubsection{Role of ionization potential}

\begin{figure}[ht]
	\centering
		\includegraphics[width=0.75\linewidth]{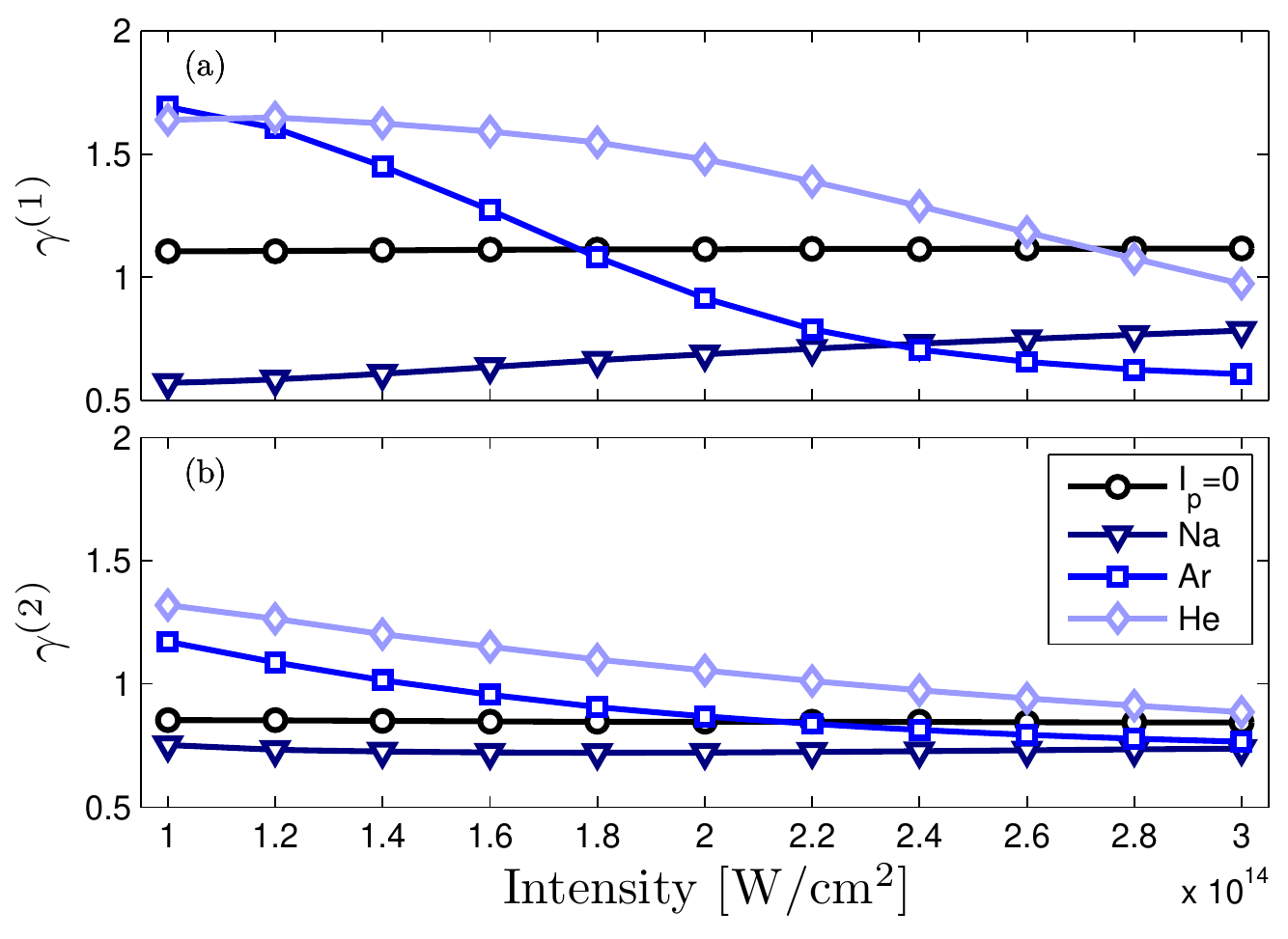}
	\caption{Ratios for different ionization potentials as a function of fundamental intensity for (a) the short branch and (b) the long branch. Symbols are $\diamond$ helium $I_p^{(He)}=24.58$ eV, $\Box$ argon $I_p^{(Ar)}=15.76$ eV, $\bigtriangledown$ sodium $I_p^{(Na)}=5.14$ eV and $\circ$ classical $I_p^{(0)}=0$ eV.}
\label{compareAtoms}
\end{figure}

In this subsection we study how $\gamma^{(n)}$, varies with laser intensity for a given laser wavelength (800 nm) and different atomic species (helium, argon, sodium and the classical case). 
In our model of $\gamma^{(n)}$, atom specific properties enter only through the value of the ionization potential, $I_p$. 
In Figure \ref{compareAtoms}, we plot $\gamma^{(n)}$ for the short branch (a) and the long branch (b). 
In the classical case ($I_p=0$) we find constant ratios:
$\gamma^{(1)}\approx 1.1$ for the short branch and $\gamma^{(2)}\approx 0.84$ for the long branch; 
while in all realistic cases, where $I_p > 0$, there is a significant change in $\gamma^{(n)}$ as a function of the fundamental intensity. 
In the case of argon there is an especially strong variation of the ratio: from 1.6 to 0.5 for the short branch over the given intensity range. This strong intensity dependence is problematic for the experimental determination of two-color HHG since the exact effective intensity is often unknown in experiments. 
%
%
%
The variation of the ratios is smaller for the long branch, because the stationary points are more similar to the corresponding classical case, as seen in Figure \ref{points}.
The variation of the ratios is smaller for sodium than for helium because of the smaller ionization potential. Sodium is chosen as an example to show that there is a significant difference between the classical and quantum mechanical cases even for a relatively small ionization potential. 
%
%
To use the two-color HHG approach for characterization of attosecond pulses would require an accurate determination of the laser intensity, as well as a quantum mechanical calculation of $\gamma^{(n)}$.
%

Next we comment on our previous work \cite{DahlstromPRA2009} where we compared the in situ method and the RABITT method for GDD of attosecond pulses. An explanation for the good agreement we found between the two methods may come from the crossing of $\gamma^{(1)}$ for argon occurring at $1.8\times10^{14}$ W/cm$^2$ with the classical limit, see Fig.~\ref{compareAtoms}(a). 
It would be interesting to see more experimental results, carried out on different atoms.
%
%
%
%

\subsubsection{Role of laser wavelength}
\begin{figure}[ht]
	\centering
		\includegraphics[width=0.75\linewidth]{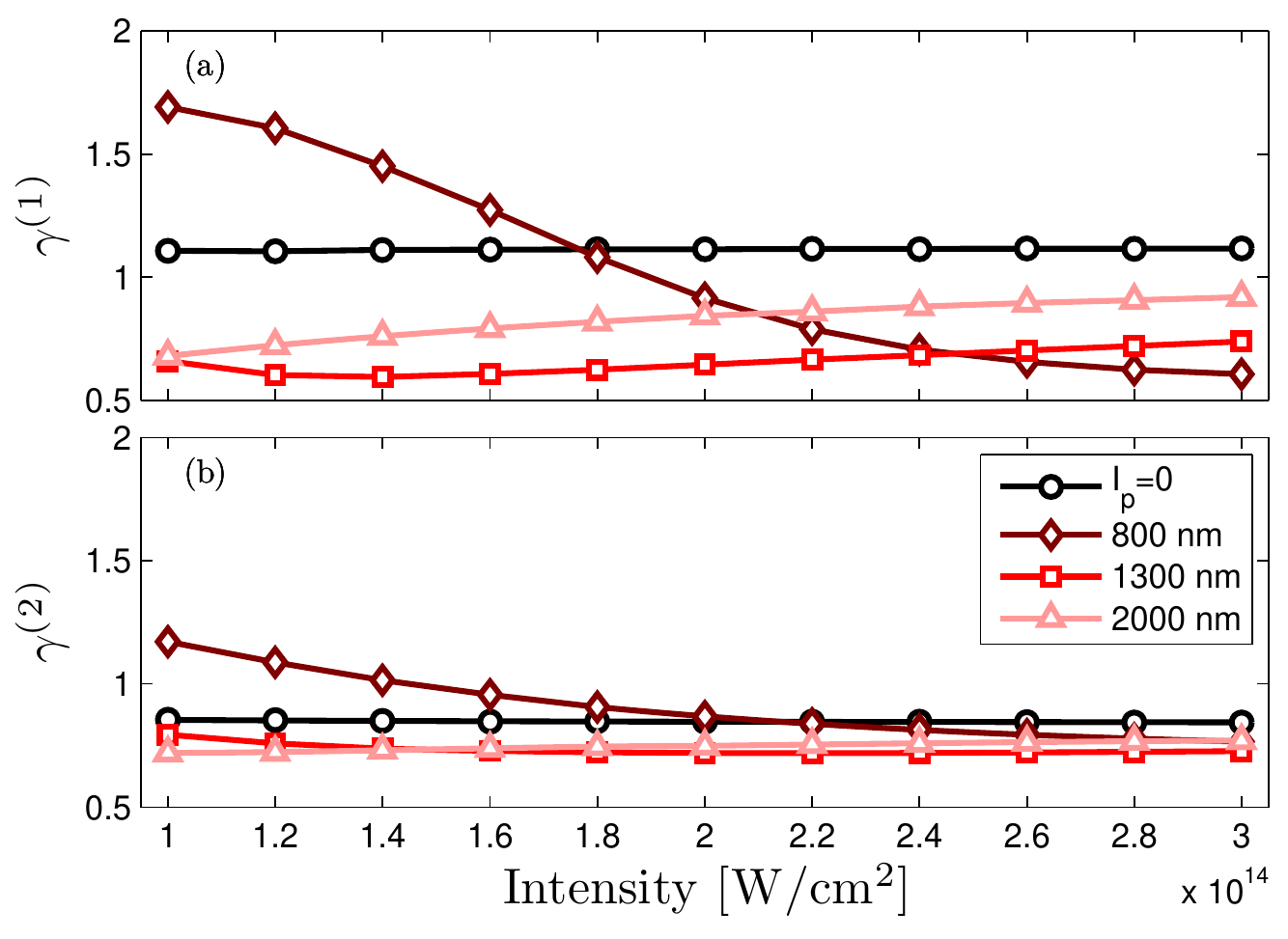}
	\caption{Ratios for different laser wavelengths as a function of fundamental intensity for (a) the short branch and (b) the long branch. Symbols are: $\diamond$ $\lambda=800$ nm, $\Box$ $\lambda=1.3$ $\mu$m, $\bigtriangleup$ $\lambda=2$ $\mu$m are for argon $I_p^{(Ar)}=15.6$ eV; while symbol $\circ$ is the classical case $I_p^{(0)}=0$ eV, which is independent of $\lambda$.}
\label{compareWavelengths}
\end{figure}

Finally, we study the dependence of the ratios, $\gamma^{(n)}$, with laser wavelength. The ratios are calculated for 800 nm, 1.3 $\mu$m and 2 $\mu$m, corresponding to a Titanium-Sapphire laser system and two mid IR laser sources. We use the ionization potential of argon. Similar to Fig.~\ref{compareAtoms}, all ratios look different when they are plotted as a function of laser intensity, see Figure~\ref{compareWavelengths} (a) for the short branch and (b) for the long branch. 
The ratios of the short branch show larger variations than those of the long branch. 
Furthermore, it is seen that the longer wavelengths lead to less variation of the ratios over intensity.
The recent experiments of Doumy et. al. \cite{DoumyPRL2009} were carried out at these wavelengths. 
Applying our method improves the scaling law for the GDD from $\lambda^{-0.77}$ to $\lambda^{-1.05}$, which is closer to the expected $\lambda^{-1}$ scaling of the harmonic chirp times intensity. 
We stress that in order to use the calculated ratios  to determine the GDD, the experimental intensity must first be measured as accurately as possible independently of the two-color HHG scheme. 

\section{Discussion and conclusions}

\begin{figure}[ht]
	\centering
		\includegraphics[width=0.75\linewidth]{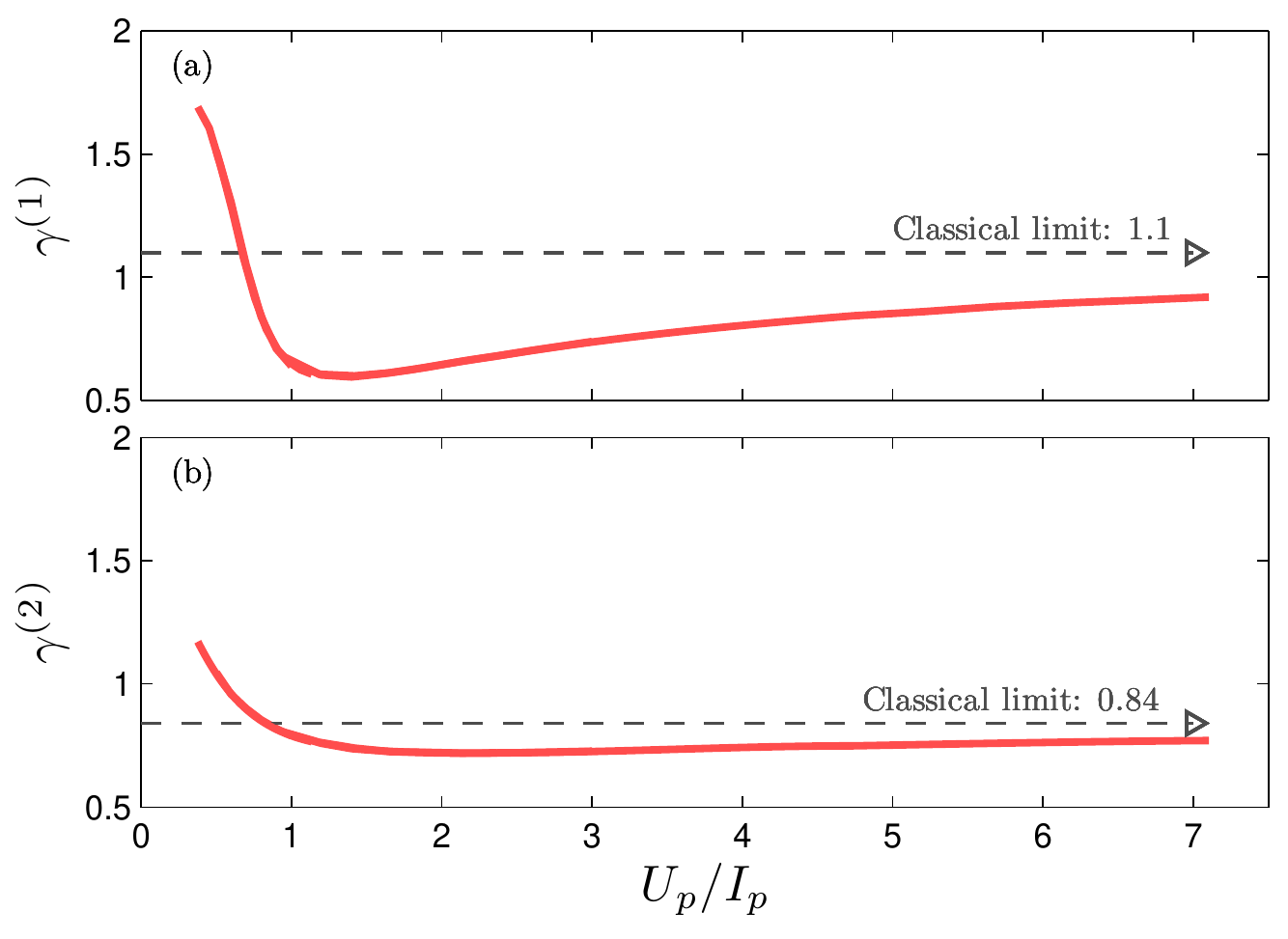}
	\caption{Universal ratio between the group delay dispersion and the derivative of the in situ phases as a function of the ratio of ponderomotive energy and ionization potential for (a) the short branch and (b) the long branch. The classical limit, $I_p^{(0)}=0$, is reached very slowly see gray arrows.}
\label{generalcurve}
\end{figure}

In our quantum mechanical derivation of two-color HHG we find that both amplitude and phase effects are important. The ratio of the ponderomotive energy and the ionization potential, $U_p/I_p$, serves as a measure on how ``classical'' or how ``quantum'' the electron trajectories are. Choosing this ratio as our x-axis for the data in Fig.~(\ref{compareAtoms})-(\ref{compareWavelengths}), we find that all individual ratios $\gamma^{(n)}$, follow a universal curve, as shown in Figure~\ref{generalcurve}. 
Photoelectron emission ranging from the photon picture to the tunneling picture is described in the theory of Keldysh \cite{KeldyshJETP1965} where the ratio $\sqrt{I_p/2U_p}\ll 1$ (corresponding to $U_p/I_p\gg 1$) implies efficient tunnel ionization. 
In this limit we find that $\gamma^{(n)}$ slowly converges toward the classical limit, $I_p=0$. 
One should look at Fig.~\ref{generalcurve} with some caution since the SFA is derived for the long wavelength limit requiring: $I_p\gg \hbar \omega$ and $U_p/I_p>1/2$. 
It is clear that the behavior of the HHG process changes dramatically around $U_p/I_p \approx 1$, \textit{i.e.} when the kinetic energy of the electron is close to the potential energy of the atom.

In conclusion, we have studied the HHG process perturbed by a weak second harmonic field within the SFA. 
We find that the dependence of the even harmonics on the subcycle delay between the two fields, can not be understood using classical theory. 
Our calculations show good agreement with experimental results \cite{HePRA2010,DahlstromPRA2009} showing the change of behavior  at high energy, explained as a change of dominant quasiclassical branch of trajectories. 
We stress that there is an intra-plateau crossing between the short and the long branch which does not coincide with the true cutoff.
Furthermore, we calculate the ratio between the GDD of the attosecond pulses and the phase variation of the even-order harmonics, $\gamma^{(n)}$, as a function of intensity, wavelength and ionization potential.
The analysis method called in situ probing of the birth of attosecond pulses \cite{DudovichNP2006} must be improved by considering the influence of the atomic properties and laser parameters, before it can be applied for quantitative experimental studies.
Using the classical analysis will only lead to a qualitative prediction for the GDD of the attosecond pulses with the correct sign.  
It will be interesting to compare the results of our quantum mechanical approach to probing the birth of attosecond pulses using a two-color field with more refined calculations.

\ack
The authors would like to thank K. Varj\'u and K.J. Schafer for helpful discussions.

\bibliographystyle{unsrt}

\begin{thebibliography}{10}

\bibitem{PaulScience2001}
P.~M. Paul, E.~S. Toma, P.~Breger, G.~Mullot, F.~Aug\'e, Ph. Balcou, H.~G.
  Muller, and P.~Agostini.
\newblock Observation of a train of tttosecond pulses from high harmonic
  generation.
\newblock {\em Science}, \textbf{292}:1689, 2001.

\bibitem{HentschelNature2001}
M.~Hentschel, R.~Kienberger, Ch. Spielmann, G.~A. Reider, N.~Milosevic,
  T.~Brabec, P.~Corkum, U.~Heinzmann\ss, M.~Drescher, and F.~Krausz.
\newblock Attosecond metrology.
\newblock {\em Nature}, \textbf{414}:509, 2001.

\bibitem{LongPRA95}
S.~Long, W.~Becker, and J.~K. McIver.
\newblock Model calculations of polarization-dependent two-color high-harmonic
  generation.
\newblock {\em Phys. Rev. A}, 52(3):2262--2278, Sep 1995.

\bibitem{GaardePRA1996}
M.~B. Gaarde, A.~L'Huillier, and M.~Lewenstein.
\newblock Theory of high-order sum and difference frequency mixing in a strong
  bichromatic laser field.
\newblock {\em Phys. {R}ev. {A}}, 54(5):4236--4248, November 1996.

\bibitem{RadnorPRA2008}
S.~B.~P. Radnor, L.~E. Chipperfield, P.~Kinsler, and G.~H.~C. New.
\newblock Carrier-wave steepened pulses and gradient-gated high-order harmonic
  generation.
\newblock {\em Phys. Rev. A}, 77(3):033806, Mar 2008.

\bibitem{ChipperfieldPRL2009}
L.~E. Chipperfield, J.~S. Robinson, J.~W.~G. Tisch, and J.~P. Marangos.
\newblock Ideal waveform to generate the maximum possible electron recollision
  energy for any given oscillation period.
\newblock {\em Phys. {R}ev. {L}ett.}, 102:063003, 2009.

\bibitem{FigueiraPRA2000}
C.~Figueira~de Morisson~Faria, D.~B. Milo\ifmmode \check{s}\else
  \v{s}\fi{}evi\ifmmode~\acute{c}\else \'{c}\fi{}, and G.~G. Paulus.
\newblock Phase-dependent effects in bichromatic high-order harmonic
  generation.
\newblock {\em Phys. Rev. A}, 61(6):063415, May 2000.

\bibitem{MauritssonPRL2006}
J.~Mauritsson, P.~Johnsson, E.~Gustafsson, A.~L'Huillier, K.~J. Schafer, and
  M.~B. Gaarde.
\newblock Attosecond {P}ulse {T}rains {G}enerated {U}sing {T}wo {C}olor {L}aser
  {F}ields.
\newblock {\em Phys. {R}ev. {L}ett.}, 97:013001, 2006.

\bibitem{ManstenNJP2008}
E.~Mansten, J.~M. Dahlstr\"om, P.~Johnsson, M.~Swoboda, A.~L'Huillier, and
  J.~Mauritsson.
\newblock Spectral shaping of attosecond pulses using two-colour laser fields.
\newblock {\em New. {J}. {P}hys.}, 10(8):083041, 2008.

\bibitem{ishiiOE2008}
N.~Ishii, A.~Kosuge, T.~Hayashi, T.~Kanai, J.~Itatani, S.~Adachi, and
  S.~Watanabe.
\newblock Quantum path selection in high-harmonic generation by a phase-locked
  two-color field.
\newblock {\em Optical {E}xpress}, 16:20876--20883, 2008.

\bibitem{dudovichPRA2009}
N.~Dudovich, J.~L. Tate, Y.~Mairesse, D.~M. Villeneuve, P.~B. Corkum, and M.~B.
  Gaarde.
\newblock Subcycle spatial mapping of recollision dynamics.
\newblock {\em Phys. {R}ev.{A}}, 80:011806(R), 2009.

\bibitem{FrolovPRA2010}
M.~V. Frolov, N.~L. Manakov, A.~A. Silaev, and N.~V. Vvedenskii.
\newblock Analytic description of high-order harmonic generation by atoms in a
  two-color laser field.
\newblock {\em Phys. Rev. A}, 81(6):063407, Jun 2010.

\bibitem{HePRA2010}
X.~He, J.~M. Dahlstr\"om, R.~Rakowski, C.~M. Heyl, A.~Persson, J.~Mauritsson,
  and A.~L'Huillier.
\newblock Interference effects in two-color high-order harmonic generation.
\newblock {\em Phys. {R}ev. {A}}, 82:033410.

\bibitem{EilanlouOX2010}
A.~Amani Eilanlou, Yasuo Nabekawa, Kenichi~L. Ishikawa, Hiroyuki Takahashi,
  Eiji~J. Takahashi, and Katsumi Midorikawa.
\newblock Frequency modulation of high-order harmonic fields with synthesis of
  two-color laser fields.
\newblock {\em Opt. Express}, 18(24):24619--24631, Nov 2010.

\bibitem{MairesseScience2003}
Y.~Mairesse, A.~de~Bohan, L.~J. Frasinski, H.~Merdji, L.~C. Dinu,
  P.~Monchicourt, P.~Breger, M.~Kova$\mathrm{\check{c}}$ev, R.~Ta\"ieb,
  B.~Carr\'e, H.~G. Muller, P.~Agostini, and P.~Sali\`eres.
\newblock Attosecond synchronization of high-harmonic soft {X}-rays.
\newblock {\em Science}, 302:1540, 2003.

\bibitem{DudovichNP2006}
N.~Dudovich, O.~Smirnova, J.~Levesque, Y.~Mairesse, M.~Yu. Ivanov, D.~M.
  Villeneuve, and P.~B. Corkum.
\newblock Measuring and controlling the birth of attosecond {XUV} pulses.
\newblock {\em Nature {P}hys.}, 2:781, 2006.

\bibitem{DoumyPRL2009}
G.~Doumy, J.~Wheeler, C.~Roedig, R.~Chirla, P.~Agostini, and L.~F. DiMauro.
\newblock Attosecond {S}ynchronization of {H}igh-{O}rder {H}armonics from
  {M}idinfrared {D}rivers.
\newblock {\em Phys. {R}ev. {L}ett.}, 102:093002, 2009.

\bibitem{DahlstromPRA2009}
J.~M. Dahlstr\"{o}m, T.~Fordell, E.~Mansten, T.~Ruchon, M.~Gisselbrecht,
  K.~Kl\"{u}nder, M.~Swoboda, A.~L'Huillier, and J.~Mauritsson.
\newblock Atomic- and macroscopic measurements of attosecond pulse trains.
\newblock {\em Phys. {R}ev. {A}}, 80:033836, 2009.

\bibitem{LewensteinPRA1995}
M.~Lewenstein, K.~C. Kulander, K.~J. Schafer, and P.~H. Bucksbaum.
\newblock Rings in above-threshold ionization: {A} quasiclassical analysis.
\newblock {\em Phys. {R}ev. {A}}, 51:1495, 1995.

\bibitem{LewensteinPRA1994}
M.~Lewenstein, Ph. Balcou, M.Yu. Ivanov, A.~L'Huillier, and P.~B. Corkum.
\newblock Theory of high-order harmonic generation by low-frequency laser
  fields.
\newblock {\em Phys. {R}ev. {A}}, 49:2117, 1994.

\bibitem{BeckerAAMO2002}
W.~Becker, F.~Grasbon, R.~Kopold, D.~B. Milosevi\'c, G.~G. Paulus, and
  H.~Walter.
\newblock Above-threshold ionization: from classical features to quantum
  effects.
\newblock {\em Advances in {A}tomic, molecular, and optical physics}, 48:35,
  2002.

\bibitem{ChirilaPRA2010}
C.~C. Chiril\ifmmode~\u{a}\else \u{a}\fi{}, I.~Dreissigacker, E.~V. van~der
  Zwan, and M.~Lein.
\newblock Emission times in high-order harmonic generation.
\newblock {\em Phys. Rev. A}, 81:033412, 2010.

\bibitem{VarjuJMO2005}
K.~Varju, Y.~Mairesse, B.~Carre, M.~B. Gaarde, P.~Johnsson, S.~Kazamias,
  R.~Lopez-Martens, J.~Mauritsson, K.~J. Schafer, Ph. Balcou, A.~L'huillier,
  and P.~Sali\`eres.
\newblock Frequency chirp of harmonic and attosecond pulses.
\newblock {\em J. {M}od. {O}pt.}, 52:379, 2005.

\bibitem{KrausePRL1992}
J.~L. Krause, K.~J. Schafer, and K.~C. Kulander.
\newblock High-order harmonic generation from atoms and ions in the high
  intensity regime.
\newblock {\em Phys. {R}ev. {L}ett.}, \textbf{68}:3535, 1992.

\bibitem{CorkumPRL1993}
P.~B. Corkum.
\newblock Plasma perspective on strong-field multiphoton ionization.
\newblock {\em Phys. {R}ev. {L}ett.}, \textbf{71}:1994, 1993.

\bibitem{KennedyPRA1972}
D.~J. Kennedy and S.~T. Manson.
\newblock Photoionization of the {N}oble {G}ases: {C}ross {S}ections and
  {A}ngular {D}istributions.
\newblock {\em Phys. {R}ev. {A}}, 5:227, 1972.

\bibitem{LewensteinPRA1995b}
M.~Lewenstein, P.~Sali\`eres, and A.~L'Huillier.
\newblock Phase of the atomic polarization in high-order harmonic generation.
\newblock {\em Phys. {R}ev. {A}}, 52:4747, 1995.

\bibitem{KeldyshJETP1965}
L~.V. Keldysh.
\newblock Ionization in the {F}ield of a {S}trong {E}lectromagnetic {W}ave.
\newblock {\em Sov. {P}hys. {JETP}}, 20:1307, 1965.

\end{thebibliography}

\hyphenation{Post-Script Sprin-ger}

\end{document}